\documentclass[prl,twocolumn,showpacs,showpacs]{revtex4}
\usepackage{graphicx}

\begin{document}

\title{Surprising lack of magnetism in the conductance channels of Pt atomic chains }
\author{Manohar Kumar}
\affiliation{Kamerlingh Onnes Laboratorium, Universiteit Leiden,
Postbus 9504, 2300 RA Leiden, The Netherlands.}
\author{Oren Tal}
\affiliation{Kamerlingh Onnes Laboratorium, Universiteit Leiden,
Postbus 9504, 2300 RA Leiden, The Netherlands.}
\affiliation{Department of Chemical Physics, Weizmann Institute of
Science, Rehovot 76100, Israel.}
\author{Roel H.M. Smit}
\affiliation{Kamerlingh Onnes Laboratorium, Universiteit Leiden,
Postbus 9504, 2300 RA Leiden, The Netherlands.}
\author{Jan M. van Ruitenbeek$^{*}$}
\affiliation{Kamerlingh Onnes Laboratorium, Universiteit Leiden,
Postbus 9504, 2300 RA Leiden, The Netherlands.}


\begin{abstract}
Pt is known to show spontaneous formation of chains of metal atoms
upon breaking a metallic contact. From model calculations these
have been predicted to be spin polarized, which is reasonable in
view of the Stoner enhanced susceptibility of bulk Pt and the
increased density of states due to the reduced dimensionality.
Here, we demonstrate that shot noise reveals information on the
magnetic state of Pt atomic chains. Against all predictions, we
find clear evidence for a non-magnetic ground state for the
conductance channels of Pt atomic chains.
\end{abstract}
\maketitle

At nanometer size scales one often finds unexpected behavior of
matter. A particularly appealing example is the spontaneous
formation of chains of metal atoms upon breaking a metallic
contact \cite{yanson98,ohnishi98}. Pt is a metal with a modestly
Stoner-enhanced magnetic susceptibility, indicating proximity to a
ferromagnetic state. A transition to ferromagnetism can be induced
by reducing dimensions, as evidenced by recent work on Pt clusters
\cite{liu06}. For these reasons, the ferromagnetic order predicted
from model calculations for atomic chains
\cite{delin03,fernandez05,smogunov08,smogunov08a,thiess09} was not
fully unexpected. Experimentally, it is very hard to design a
probe that can directly measure the magnetism of atomic chains.
Here, we demonstrate that shot noise, the intrinsic noise due to
the discrete character of the electronic charge, reveals
information on the magnetic state of Pt atomic chains. We find
clear evidence for a non-magnetic ground state of the conductance
channels for Pt atomic chains.

Shot noise was first discussed for vacuum diodes by Schottky
\cite{schottky18}, who showed that this current noise is
independent of frequency (white noise) up to very high
frequencies, and its power spectrum has a value of $S_I=
2e\overline{I}$, with $e$ the absolute value of the electron
charge, and $\overline{I}$ the average current. In nanoscale
conductors, for which the system size is much smaller than the
electron scattering length, this noise can be understood as
partition noise. In these systems the number of transmission
channels available for electrons to cross a conductor is limited
and the transmission through each one of the channels is set by
the properties of the conductor. When the transmission probability
is smaller than 1 the conductor can be viewed as an effective
bottleneck causing a random sequence of electron backscattering
events, which is observed as current fluctuations or noise. The
theory has been elaborated by several groups and has been
thoroughly reviewed by Blanter and B{\"u}ttiker \cite{blanter00}.
For a nanoscale conductor with $N$ conductance channels, each
characterized by a transmission probability $\tau_{n}$, the
current noise power at an applied bias voltage $V$ is given by,
\begin{equation}
S_I = 2eV \coth \left( \frac{eV}{2k_{\rm B} T} \right)
\frac{e^2}{h} \sum_{n=1}^N \tau_{n} (1-\tau_{n})
+ ~ 4k_{\rm B} T ~ \frac{e^2}{h}  \sum_{n=1}^N \tau_n ^2 ~.
\label{eq:coth}
\end{equation}
\noindent where $k_{\rm B}$ is Boltzmann's constant, and $T$ is
the temperature of the nanoscale conductor. Anticipating spin
splitting of the conductance channels we treat conductance
channels for each spin direction separately. In equilibrium (at
$V=0$) equation~(\ref{eq:coth}) reduces to the Johnson-Nyquist
thermal noise, $4k_{\rm B} TG$, describing the current
fluctuations that are driven only by the thermal motion of
electrons. $G=(e^2/h)\sum \tau_n$ is the conductance. Again, in
the expression for $G$ we take the conductance quantum as $e^2/h$
and sum over spin states. In the low-temperature limit, $k_{\rm B}
T \ll eV$, equation~(\ref{eq:coth}) reduces to $S_I=
2e\overline{I}F$, where the Fano factor $F$ measures the quantum
suppression of Schottky's classical result,
\begin{equation}
F=\frac{\sum_n \tau_n (1-\tau_n)}{\sum_n \tau_n}.
\end{equation}
From this analysis it is apparent that one may obtain information
on the transmission probabilities of the conductance channels by
measurement of the noise power, and in favorable cases it is even
possible to determine the number of conductance channels
\cite{brom99}. The Fano factor reduces to zero when all
conductance channels are either fully blocked ($\tau_n = 0$), or
fully open ($\tau_n =1$). For a nanowire with a given conductance
$G=(e^2/h)\sum \tau_n$ the noise has a lower bound that is
obtained by taking all open channels to have perfect transmission,
except for one that takes the remaining fraction of the
conductance. This minimum will sensitively depend on whether the
spin channels are restricted to be degenerate. It is this property
that we exploit when investigating the magnetic state of Pt atomic
chains.

\begin{figure}[t!]
\includegraphics[width=8cm]{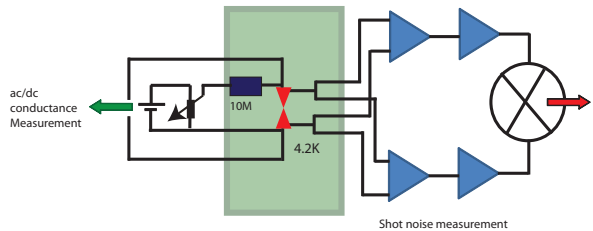}
\caption{(Color online) Schematics of the circuit used for
measuring conductance and noise on an atomic junction formed by
the MCBJ technique. Noise is measured by using two sets of
low-noise amplifiers, with a total amplification factor of $10^5$,
and by taking the cross spectrum of the two channels in a
frequency range between 250Hz and 100kHz. After averaging of
$10^4$ spectra the uncorrelated noise of the preamplifiers is
strongly suppressed.} \label{fig:circuit}
\end{figure}

Platinum atomic junctions were formed at liquid helium
temperatures using mechanically controllable break junctions
(MCBJ, for more details see Refs.~\cite{djukic06,agrait03,suppl}.
The electronic circuit for the measurement is shown schematically
in Fig~\ref{fig:circuit}. The Pt contact was first characterized
by recording a conductance histogram \cite{smit01,suppl}. The
conductance histogram shows a first peak at a conductance of about
$1.5\cdot(2e^2/h$) with very few conductance counts below
$1\cdot(2e^2/h)$, as expected for clean Pt point contacts: Pt
being an s-d metal has up to 12 conductance channels due to the
six s and d orbitals, and spin. Each of the channels has a finite
transmission probability and they sum up to a total of about
$1.5\cdot(2e^2/h$), in agreement with calculations
\cite{nielsen02,vega04,fernandez05,smogunov08}. The strong peak at
$1.5\cdot(2e^2/h$) reflects the frequent formation of atomic
chains in the contact. Chain formation can be demonstrated more
explicitly by recording a histogram of the length of the
conductance plateaux with conductances in the range of the first
conductance peak, between 1.2 and 2 times ($2e^2/h$)
\cite{yanson98,smit01,untiedt02,suppl}.

\begin{figure}[t!]
\includegraphics[width=6cm]{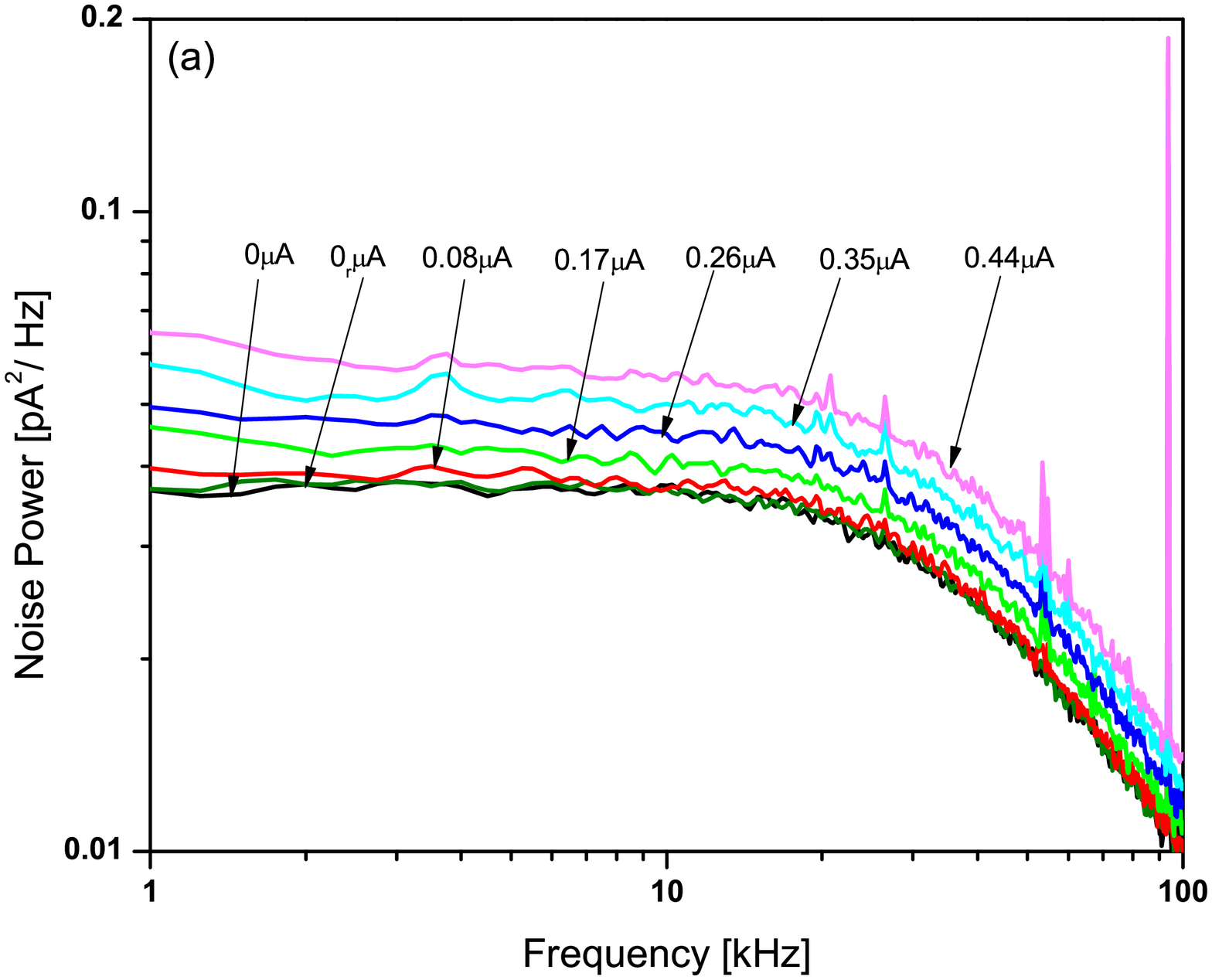}
\includegraphics[width=7cm]{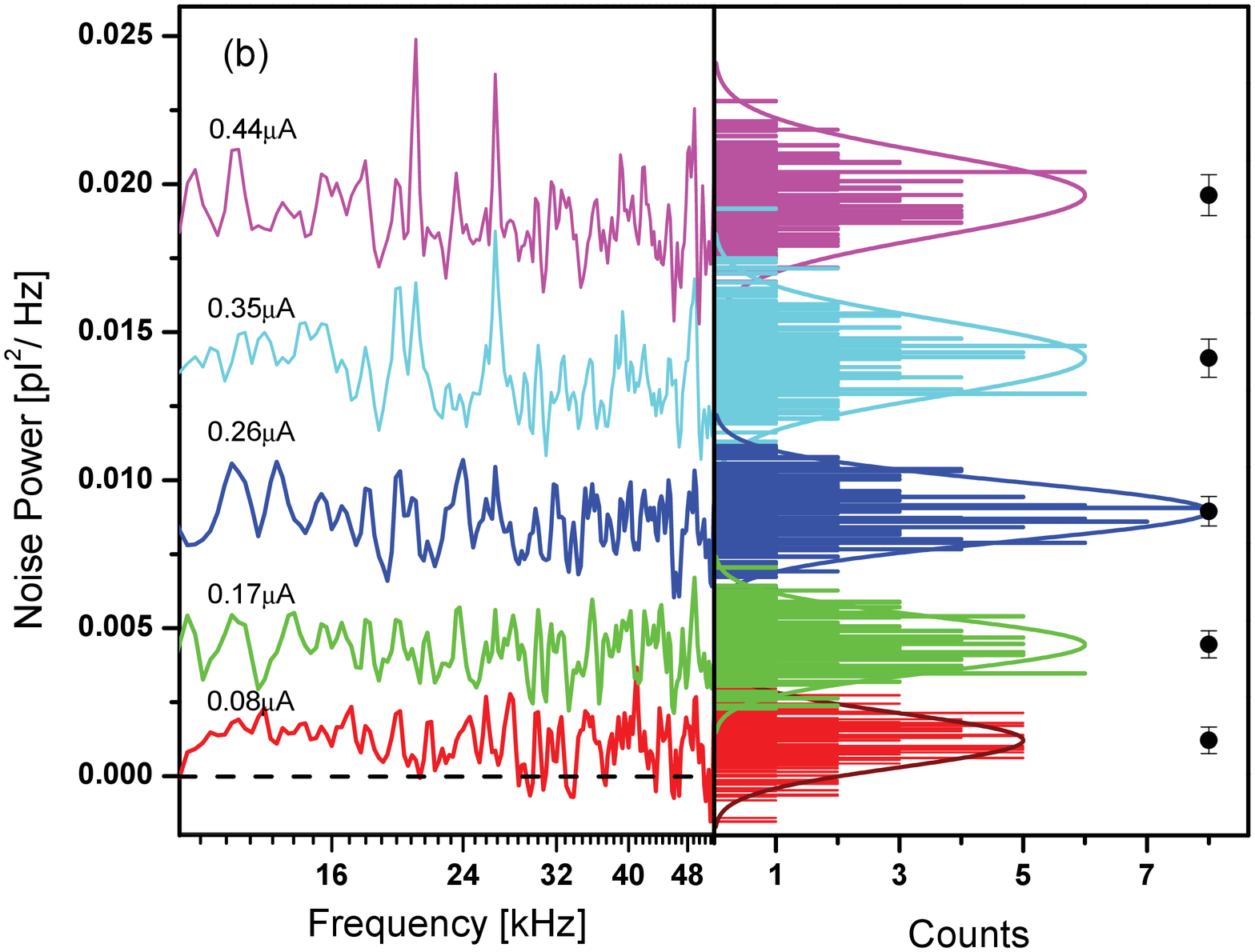}
\caption{ Example of the noise power data analysis. (a) Noise
power spectra for a Pt atomic chain of $\sim 3$ atoms in length
having a conductance $G=1.52\cdot(2e^2/h)$ and a Fano factor
F=0.192. The peaks are due to spurious signals that could not be
fully shielded. (b) Same data after subtracting the thermal noise
and correcting for the roll-off. The spurious signals are
effectively removed by the subtraction procedure. }
\label{fig:raw_data}
\end{figure}

After this preliminary characterization of the junction an atomic
chain was made by pulling, starting from a large contact until the
conductance drops to a value near $1.8\cdot(2e^2/h$). Measurements
of conductance and noise were taken at several points of
subsequent stretching starting from here. The corresponding piezo
voltages were recorded in order to identify the length in terms of
the mean number of atoms forming the chain. The zero-bias
differential conductance, $dI/dV$, was recorded, which is needed
in combination with the noise for the analysis of the conductance
channels. The accuracy of the ac conductance measurement is better
than 1\%, as verified by tests on standard resistors.
 Fig.~\ref{fig:raw_data} shows an example of
noise spectra taken in the window from 1kHz to 100kHz for a series
of current settings, and illustrates how the noise power is
obtained from the data. First, the thermal noise is recorded at
zero bias, and after taking noise spectra at several bias settings
the zero bias noise is recorded once more (labelled as $0_r
\rm{\mu A}$)  in order to verify that the junction has remained
stable. The low-frequency upturn at larger currents is due to
1/f-like noise. At high frequencies there is a roll-off due to the
transfer characteristics of the circuit. The thermal noise level
corresponds to a temperature of 6.3 K, which agrees within the
accuracy of the temperature measurement with a reading of 6.1 K,
as obtained from a ruthenium oxide 10k resistance thermometer. For
several junction settings conductance measurements were repeated
after the shot noise bias sequence in order to detect possible
changes in the conductance. Typical changes observed were smaller
than 2\%. Fig.~\ref{fig:raw_data}(b) shows that the spectra become
white above 10kHz after correction for the roll-off with a single
RC time constant. The thermal noise (at zero bias) is subtracted,
which explains the negative values in the data fluctuations for
the lowest currents. The data points are projected in the form of
a histogram, shown at the right, and the level of white noise is
obtained from the center of the histogram for each voltage bias.
The bullets and error bars at the right indicate the position and
accuracy of the noise power as determined from a gaussian fit to
the histograms.

Since shot noise and thermal noise are of comparable magnitude in
these experiments it is useful to represent the data such that the
expected dependence on the applied bias in Eq.~(\ref{eq:coth}) is
apparent. The voltage dependence in Eq.~(\ref{eq:coth}) can be
lumped into a single variable $X$ that we take to be $X=x\coth x$,
with $x=eV/2k_{\rm B}T$. The reduced excess noise is then defined
as,
\begin{equation}
Y=\frac{S_I (V) -S_I (0)}{S_I (0)},
\end{equation}
where $S_I (V)$ is the noise at finite bias, and $S_I (0)$ is the
thermal noise, at zero bias. The reduced excess noise is now
expected to depend linearly on the control parameter, $Y=(X-1)F$,
from which the Fano factor $F$ can be easily obtained.

\begin{figure}[b!]
\includegraphics[width=8cm]{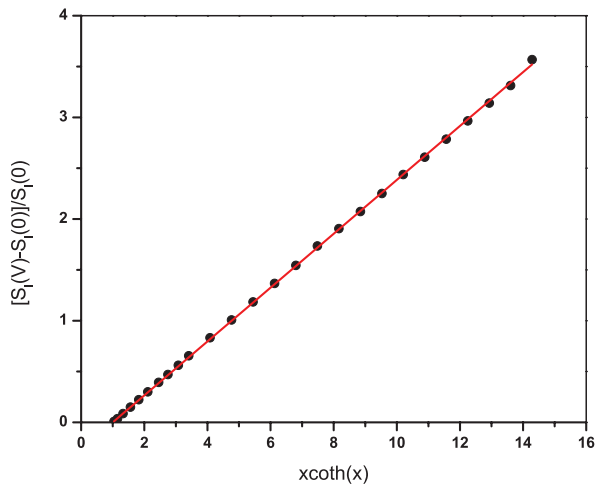}
\caption{Reduced excess noise $Y=(S_I (V) -S_I(0))/S_I (0)$ for a
Pt atomic chain. The excess noise is given as a function of
$X=x\coth (x)=(eV/2k_{\rm B} T)\coth (eV/2k_{\rm B} T)$, for a
chain having a conductance of $G=1.425\pm 0.01(2e^2/h)$ at a
length of about 2 atoms in the chain.  } \label{fig:reduced-plot}
\end{figure}

Figure~\ref{fig:reduced-plot} shows a series of measurements on a
Pt atomic chain with a conductance of $G=1.425\pm 0.01(2e^2/h)$ at
a short length of 2 atoms in the chain, for 26 settings of the
bias voltage in the range from  0mV to 16.6mV (0 to 1.83$\rm{\mu
A}$). The slope of the plot gives a Fano factor $F=0.269\pm
0.009$. The accuracy for each of the points is 3\%, as obtained by
a fit to the power spectrum after correction for the roll-off as
in Fig.~\ref{fig:raw_data}. The measurement required about 50
minutes, illustrating the long-term stability of the atomic
chains. It shows a very nice agreement with the expected
dependence, and the scatter around the linear slope is within the
data point accuracy.

\begin{figure}[t!]
\includegraphics[width=8.7cm]{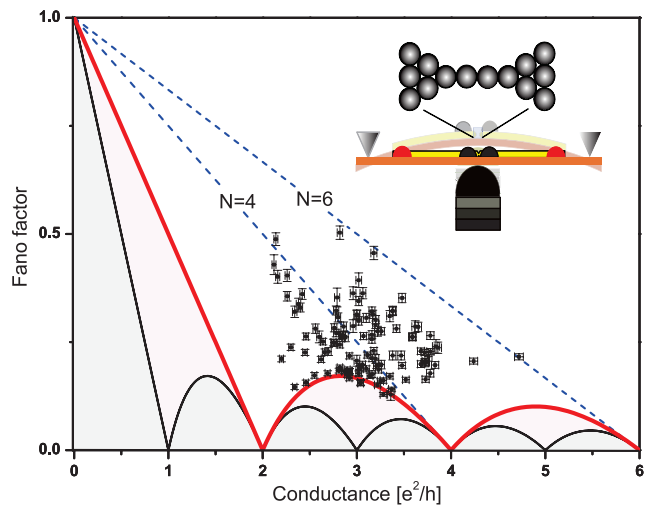}
\caption{Fano factor vs. conductance for 119 different Pt atomic
chain configurations. The bold red curve shows the minimum noise
curve when spin degeneracy is imposed. Relaxing spin degeneracy
results in a minimum noise curve shown by the thin black curve.
The inset illustrates the principle of the break junction
experiment.} \label{fig:Fano-vs-G}
\end{figure}

We have recorded similar plots for over 500 configurations of Pt
atomic chains of various length, for which we took 7 bias voltage
points between 0 and 0.44$\rm{\mu}$A. When the scatter in the plot
of the reduced excess noise was larger than 3\%, or the thermal
noise at start and end of the measurement differed by more than
2\%, we rejected the data. The scatter is mostly due to a large
$1/f$ component in the noise spectrum and the contribution of the
residual amplifier noise correlations to the spectra. After this
selection 119 configurations remain. Figure~\ref{fig:Fano-vs-G}
shows the Fano factors determined from these 119 sets of shot
noise measurements.

The bold red curve shows the minimum noise curve when spin
degeneracy is imposed. Relaxing spin degeneracy results in a
minimum noise curve shown by the thin black curve. The blue broken
lines show the {\em maximum} noise that can be obtained with $N=4$
or $N=6$ (spin) channels. This maximum is obtained by taking all
channels to have the same transmission probability $\tau =
Gh/e^2N$, leading to $F=1-\tau$. The measured data points form a
diffuse cloud in $(G,F)$-space, with its centre of mass near
$G=1.5\cdot(2e^2/h)$ and a spread in the conductance in agreement
with the position and width of the first peak in the conductance
histogram. A large fraction of the points lie above the line
labelled $N=4$, which represents the maximum Fano factor when only
four channels are available. This shows that these Pt atomic
chains have at least five conductance channels, in agreement with
calculations \cite{nielsen02,fernandez05,smogunov08}. The points
below the blue broken line $N=4$ can be explained by four
channels, but for the majority of points the only conclusion we
can draw is that {\em at least} four channels are involved.

The most striking observation, and the central point of this
Letter, is that all Fano factors for the Pt chain configurations
fall on, or well above, the curve describing the minimum noise for
{\em spin-degenerate} channels.  More than 15\% of the measured
points are even found to coincide within the error bars with the
minimum-Fano curve for spin-degenerate channels, and none of the
points are found significantly below it. For spin-split
conductance channels the limiting curve is represented by the thin
curve in Fig.~\ref{fig:Fano-vs-G} \cite{roche04,dicarlo06}. This
provides strong evidence that the conductance channels in the Pt
atomic chains formed in the experiment are spin degenerate, at
variance with results from DFT calculations
\cite{delin03,fernandez05,smogunov08,smogunov08a,thiess09}. Most
points in Fig.~\ref{fig:Fano-vs-G} are measured for Pt chains of 3
to 4 atoms in length, occasionally 5 or 6 atoms. We do not find
any systematic evolution of the Fano factor with stretching of the
chain. While increasing the length of the chain in steps the Fano
factor may be seen to jump towards the minimum noise curve, but
then it jumps away from the curve to higher values at next steps
in increasing the chain length (see \cite{suppl}).

A second remarkable observation is the fact that there is a group
of 18 points that coincide with the curve describing the minimum
noise power for spin-degenerate channels. This is quite
unexpected. Whether spin degenerate or spin polarized, all
calculations predict that at least six spin channels are involved,
and no mechanism is known that would lead to unit transmission for
the dominant channels
\cite{nielsen02,vega04,fernandez05,smogunov08}. The most complete
calculations \cite{smogunov08}, fully relativistic Density
Functional Theory for realistic chain sizes, show that a magnetic
moment above 0.4${\rm{\mu}}$B per atom already appears for a chain
only three atoms in length, at equilibrium interatomic distance.

It has been argued \cite{calvo09,untiedt10} that the zero bias
anomalies observed in the differential conductance for Pt atomic
chains provide evidence for local magnetic order. Indeed, the
differential conductance often shows a pronounced structure near
$V=0$ (see \cite{suppl}. However, this structure is very irregular
and may have any sign or structure, which hampers a
straight-forward interpretation. If any magnetic order is present
for a consistent interpretation we would have to assume such
moments would be formed by states that do not participate in the
conductance.

In conclusion, we find strong evidence for an absence of magnetic
order in the conductance channels for Pt atomic chains. The fact
that this observation disagrees with many DFT-based computations
suggests that effects beyond the present models, such as
electron-electron correlations, may need to be considered.

\acknowledgments{This work is part of the research programme of
the Foundation for Fundamental Research on Matter (FOM), which is
financially supported by the Netherlands Organisation for
Scientific Research (NWO). }

\end{document}


\title{Surprising lack of magnetism in the conductance channels of Pt atomic chains\\ Supplementary Information}
\author{Manohar Kumar}
\affiliation{Kamerlingh Onnes Laboratorium, Universiteit Leiden,
Postbus 9504, 2300 RA Leiden, The Netherlands.}
\author{Oren Tal}
\affiliation{Kamerlingh Onnes Laboratorium, Universiteit Leiden,
Postbus 9504, 2300 RA Leiden, The Netherlands.}
\affiliation{Department of Chemical Physics, Weizmann Institute of
Science, Rehovot 76100, Israel.}
\author{Roel H.M. Smit}
\affiliation{Kamerlingh Onnes Laboratorium, Universiteit Leiden,
Postbus 9504, 2300 RA Leiden, The Netherlands.}
\author{Jan M. van Ruitenbeek$^{*}$}
\affiliation{Kamerlingh Onnes Laboratorium, Universiteit Leiden,
Postbus 9504, 2300 RA Leiden, The Netherlands.}


\maketitle

\section{Experimental procedure}

Platinum atomic junctions were formed at liquid helium
temperatures using mechanically controllable break junctions
(MCBJ). The sample chamber was pumped to $\sim 10^{-5}$mbar before
cooling down in liquid helium. The chamber was fitted with active
charcoal for cryogenic pumping such that the pressure in the
chamber drops below measurable values  at liquid He temperature.
Once cold and under vacuum the Pt sample wire was first broken by
mechanical bending of the substrate. By relaxing the bending the
broken wire ends can be rejoined and the size of the contact can
be adjusted with sub-atomic precision by means of a piezo-electric
actuator.

Conductance was measured dc, or ac by means of a small modulation
voltage (2mV and 2.3kHz) and a lock-in amplifier. This circuit can
be decoupled from the sample during shot noise measurement using a
switch at room temperature. Noise was measured by using two sets
of low-noise amplifiers, with a total amplification factor of
$10^5$, and by taking the cross spectrum of the two channels in a
frequency range between 250Hz and 100kHz. After averaging of
$10^4$ spectra the uncorrelated noise of the preamplifiers is
strongly suppressed. The cryostat along with all amplifiers for
both conductance measurement and shot noise measurement were
placed in a Faraday cage that also provides acoustic shielding.
During the shot noise measurements the conductance circuit was
disconnected in order to eliminate external noise sources.

The noise spectra were recorded for a window from 250Hz to 100kHz.
An example of such spectra is given in Fig.~2 of the main text. At
the low-frequency end of the spectrum one observes an increase in
the spectrum above the white noise level due to a $1/f$-like noise
contribution, the amplitude of which varies between different
junction settings, which has been attributed to defect
fluctuations in the leads \cite{ralph92}. This part of the
spectrum is ignored for the analysis, but it influences the
accuracy of the determination of the white noise power. At the
high-frequency end of the spectrum a roll-off is seen, with a
characteristic frequency of about 50kHz that is due to the RC time
constant of the stray capacitance of the leads in combination with
the junction resistance. Finally, a slight upturn in the spectra
at the highest frequencies is due to residual correlations in the
noise of the two amplifiers. The analysis presented below
eliminates the roll-off, but is sensitive to the residual
correlations and the $1/f$ noise, which limit the accuracy in
determination of the Fano factor.

\section{Conductance histogram}

\begin{figure}[t!]
\includegraphics[width=12cm]{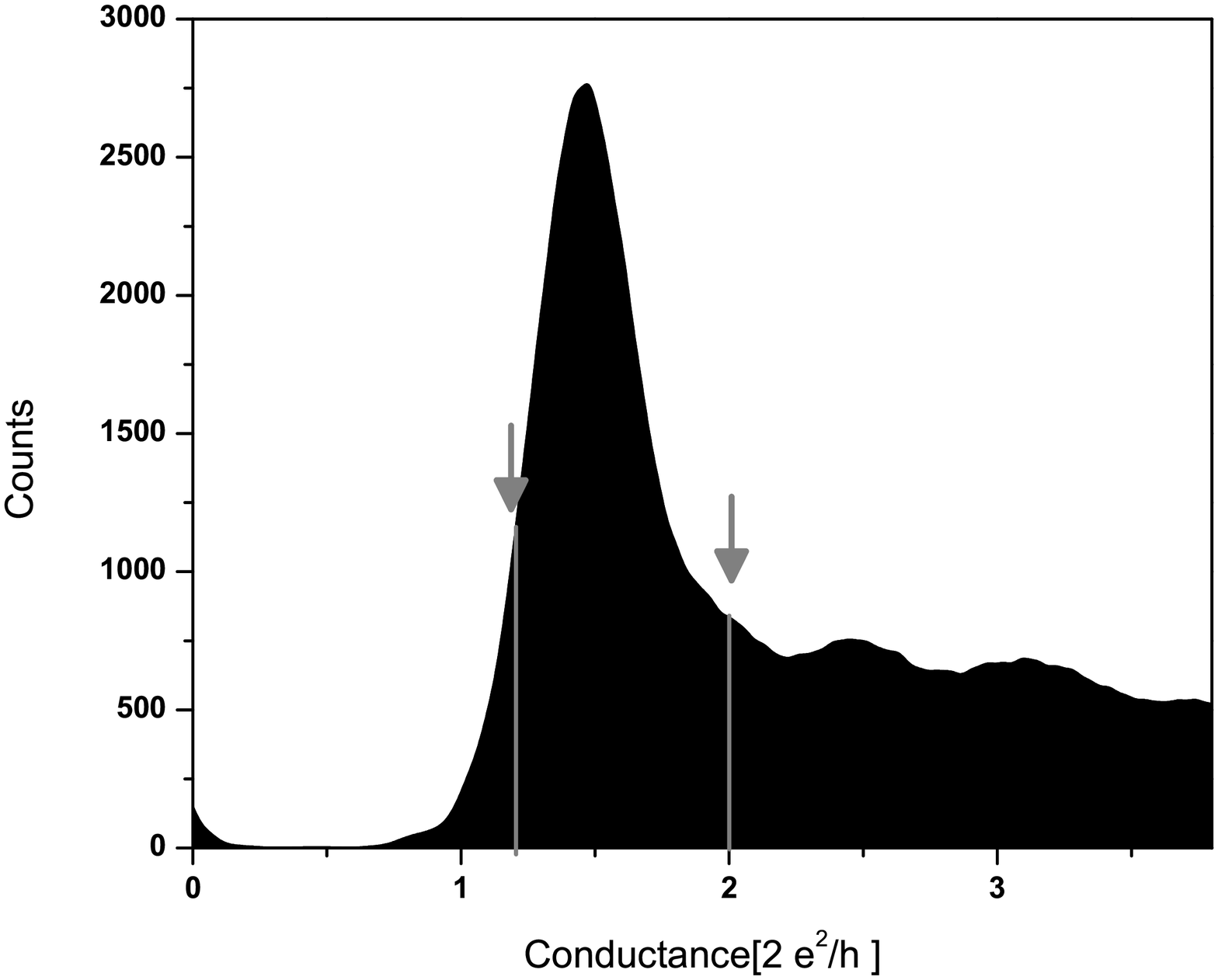}
\caption{Conductance histogram for a clean Pt contact at liquid
helium temperatures. } \label{fig:suppl-cond-histo}
\end{figure}

Before starting shot noise measurements the Pt contact was first
characterized by recording a conductance histogram,
Fig.~S\ref{fig:suppl-cond-histo}. The conductance histogram for a
clean Pt contact at liquid helium temperatures is recorded by
combining one thousand conductance breaking traces. Contacts are
repeatedly made and broken, controlled by the piezo voltage that
regulates the substrate bending of the mechanically controllable
break junction device, at a fixed bias voltage setting of 80mV.
The points of the digitized traces of conductance are collected
into a histogram and the counts are plotted as a function of the
conductance. The first peak at $\sim 1.5\cdot(2e^2/h)$ represents
the average conductance of a contact of a single Pt atom in cross
section. Below this peak the count drops to very low numbers,
indicating that the contact finally breaks to a clean vacuum
tunnel junction. Note that we use units of  $2e^2/h$ here for
reasons of comparing to earlier work, but that the conductance
units per spin, $e^2/h$, are used throughout the text. The arrows
indicate the boundaries used for recording length histograms as in
Fig.~S\ref{fig:suppl-length-histo}

\section{Length histogram}

\begin{figure}[t!]
\includegraphics[width=12cm]{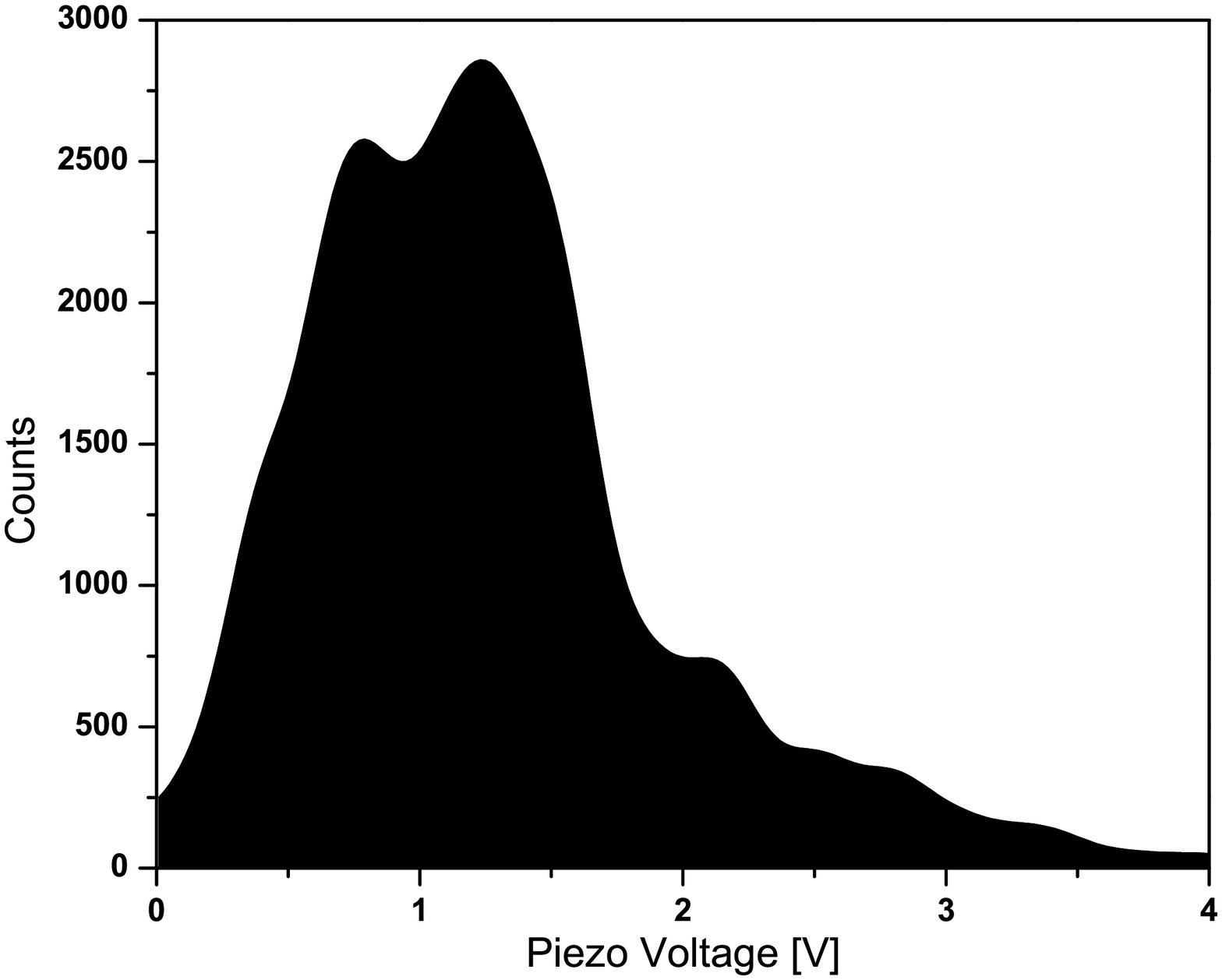}
\caption{ Length histogram recorded for a Pt break junction.  }
\label{fig:suppl-length-histo}
\end{figure}

Figure~S\ref{fig:suppl-length-histo} shows a length histogram
obtained for a clean Pt junction at low temperatures. This is used
to verify chain formation and for calibrating the displacement.
The histogram in Fig~S\ref{fig:suppl-length-histo} is obtained by
combining 4500 traces and recording the length of the conductance
plateaux with conductances between 1.2 and 2 times ($2e^2/h$),
i.e. in the range of the first conductance peak in
Fig.~S\ref{fig:suppl-cond-histo}. The length axis is given in
units of the voltage on the piezo element, where the
proportionality constant is 0.25 V/\AA. The histogram is
consistent with the earlier work of Untiedt {\it et al.}
\cite{untiedt02}. The first three peaks can be interpreted as the
lengths corresponding to chains of 2, 3, and 4 atoms.

\section{Stretching sequences of the Fano factor}

\begin{figure}[t!]
\includegraphics[width=12cm]{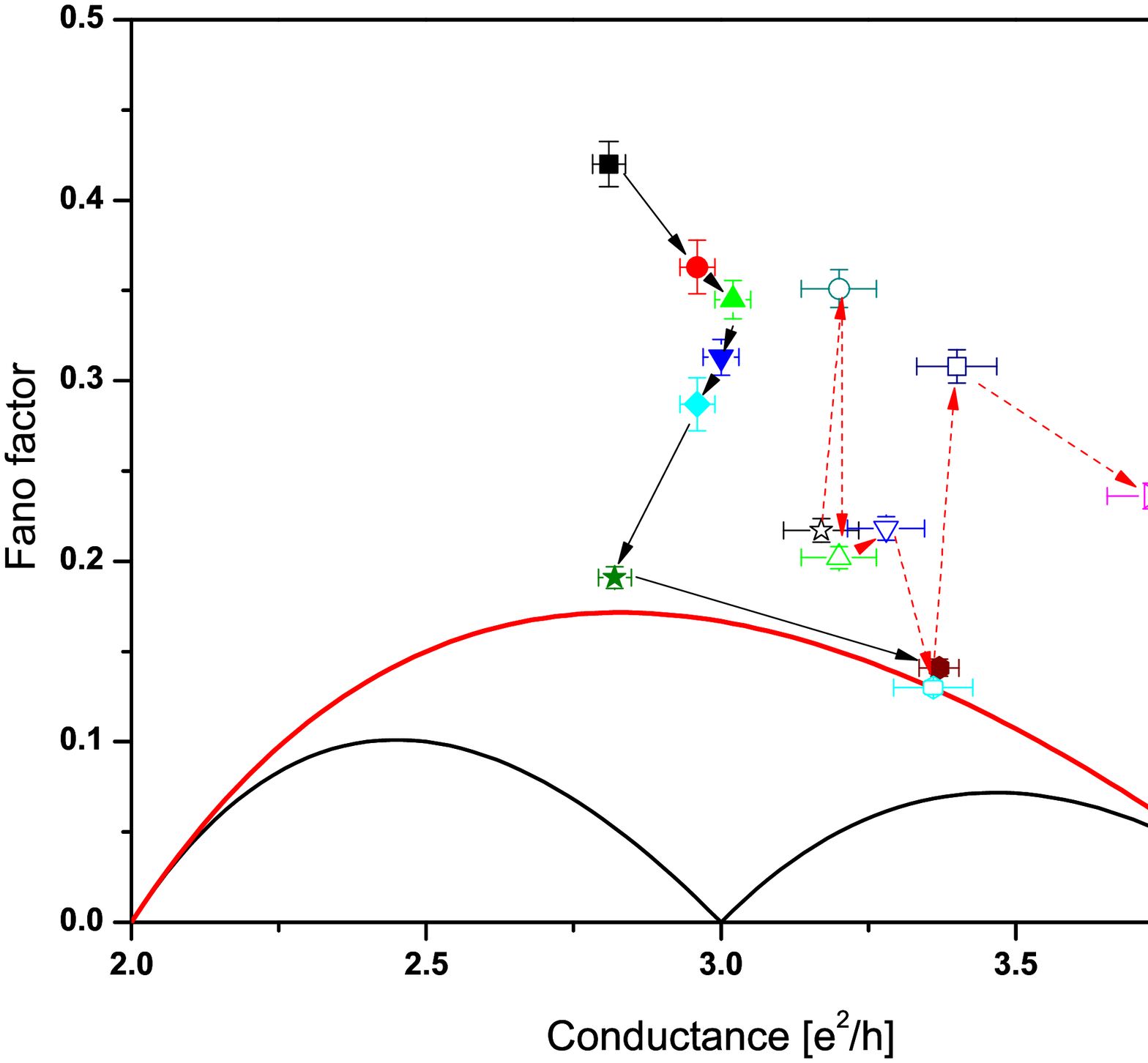}
\caption{Examples of the evolution of conductance $G$ and Fano
factor $F$ when a Pt atomic chain is stepwise elongated.  }
\label{fig:suppl-Fano-vs-stretching}
\end{figure}

Figure~S\ref{fig:suppl-Fano-vs-stretching} shows two examples of
the evolution of conductance $G$ and Fano factor $F$ when a Pt
atomic chain is stepwise elongated. The stretching sequence has a
length of 3.0V ($\sim 12$\AA, solid arrows and filled symbols) and
3.52V ($\sim 14$\AA, dashed arrows and open symbols),
corresponding to 5 or 6 atoms final length, respectively. These
two examples are chosen to illustrate that there does not appear
to be a systematic evolution towards the minimum noise curve for
longer chains, although points at the curve are more frequently
found for longer chains.

\section{Zero-bias anomaly}
Figure~S5 shows four examples of differential conductance
characteristics of Pt atomic chain junctions. A zero-bias anomaly
is typically observed, that resembles a Kondo anomaly. This
feature has been remarked before, but an unambiguous
identification of its origin was not made. It does not show a
regular peak structure expected for a Kondo anomaly, and its shape
varies strongly between chain configurations, appearing sometimes
as a peak, sometimes as a dip. Moreover, similar structure is also
found for single-atom contacts and even for much larger contacts.
Zero-bias anomalies may arise from various excitations, including
two-level atomic configurations and low-lying vibrational modes.
While a magnetic origin cannot be excluded, the related magnetic
moment does not appear to be associated with the conductance
channels of the atomic chain, as evidenced by the shot noise
measurements presented in this work.

\begin{figure}[t!]
\includegraphics[width=12cm]{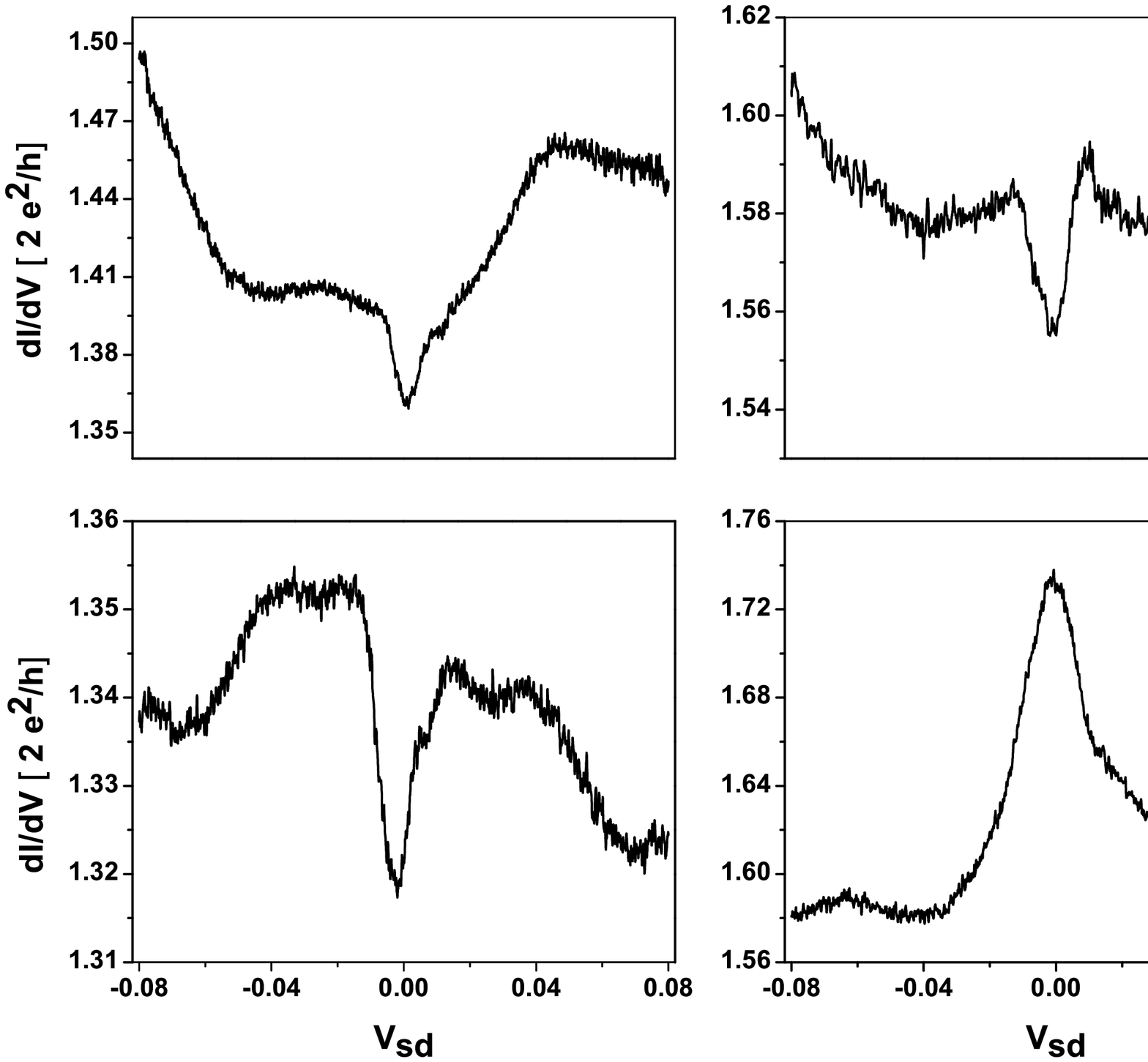}
\caption{Four examples of differential conductance characteristics
of Pt atomic chain junctions. } \label{fig:suppl-dIdV}
\end{figure}